\documentclass[a4paper,20pt]{article}
    \usepackage[T1]{fontenc}
    \usepackage{graphicx}
    \usepackage{epsfig}
    \usepackage{amsmath}
    \usepackage{amsfonts}
    \renewcommand{\abstract}{}
\begin{document}
\makeatletter
\renewcommand{\@oddhead}{\textit{YSC'14 Proceedings of Contributed Papers} \hfil \textit{R. Stevenson}}
\renewcommand{\@evenfoot}{\hfil \thepage \hfil}
\renewcommand{\@oddfoot}{\hfil \thepage \hfil}
\fontsize{11}{11} \selectfont

\title{Detection of Small Kuiper Belt Objects by Stellar Occultations}
\author{\textsl{R. Stevenson}}
\date{}
\maketitle
\begin{center} {\small Institute for Astronomy,University of Hawaii, 2680 Woodlawn Drive, Honolulu, HI 96822, USA\\
rach@ifa.hawaii.edu}
\end{center}

\begin{abstract}
Knowledge of the Kuiper Belt is currently limited to those objects
that can be detected directly. Objects with diameters less than
$\sim$10km reflect too little light to be detected.  These smaller
bodies could contain most of the mass in the Kuiper Belt while the
abundance of these bodies may constrain the distribution of mass.
The overall size distribution of bodies within the Kuiper Belt can
also be inferred from the relative abundances of sub-km and larger
bodies.

Stellar occultations are already used to study dark objects in the
Solar System, such as asteroids or planetary rings. Occultation by a
KBO of a size comparable to, or larger than, that of the Fresnel
Scale will result in Fresnel diffraction. Detection of diffraction
effects requires fast multiple-star photometry, which will be
conducted in July 2007 using the Orthogonal Parallel Transfer
Imaging Camera (OPTIC) mounted on the University of Hawaii 2.2m
telescope on Mauna Kea.  This paper details how knowledge of the
mass and structure of the outer Solar System may be obtained through
the detection of serendipitous stellar occultations.
\end{abstract}

\section*{Introduction}
\indent \indent The Kuiper Belt is the remnant of the
proto-planetary disk from which the planets formed.  Kuiper Belt
Objects (KBOs) are some of the most primordial objects in the Solar
System. The total mass, composition and spatial and size
distributions of KBOs are indicators of conditions in the early
Solar System, and the nature of subsequent processing of material.
Current knowledge is limited to objects larger than $\sim$10km that
represent only a fraction of the total population.

Over 1000 trans-Neptunian objects have been discovered since 1992.
An extrapolation of the size distribution observed for larger KBOs
suggests the Kuiper Belt has a mass of 0.1$M_{E}$ \cite{jewitt}. The
actual mass of the Kuiper Belt may be several times this value when
the smallest KBOs are taken into account.

Stellar occultations provide an indirect method to constrain the
abundance of small objects in the outer Solar System. The method
consists of monitoring the flux from a suitable target star using
fast photometry and identifying significant dips in flux. The method
has been previously used to detect rings around the icy giant
planets and to investigate planetary atmospheres (e.g.
\cite{elliot}). Bailey \cite{bailey} first suggested that stellar
occultations could detect small objects in the Solar System.  In the
last 5 years the necessary technology has become available to
acquire rapid photometry. Reported observations to date have been
sparse, with less than 100 possible detections reported \cite{Chang,
roques2}.  A larger sample is needed to reliably estimate the number
of small ($<$10km) KBOs.

Here I discuss the importance of using stellar occultations to
detect km and sub-km sized bodies in the outer Solar System.

\section*{Size Distributions within the Kuiper Belt}
\indent \indent A power law can be used to describe the size
distribution of objects in the Kuiper Belt:
\begin{equation}
N(r)=Cr^{-q}dr
\end{equation}
Empirically, the parameter q is equal to 4 for observed KBOs with
sizes greater than 10km \cite{luu}. Dynamical simulations suggest
q$\sim$3.5 for objects smaller than 10km (e.g. \cite{dohnanyi}). The
sizes of smaller bodies are primarily governed by collisional
processes while sizes of larger objects are controlled by
accretional processes as the gravitational influence of the bodies
becomes important.

Jupiter Family Comets (JFCs) are probably the km-sized progeny of
KBOs that were recently scattered into the inner Solar System. There
is some evidence that the size distribution governing JFCs is
similar to that governing the smaller KBOs \cite{tancredi}, which
would support the theory that JFCs originate from the Kuiper Belt.

Fig. 1 shows the size distribution over time as modeled by Kenyon \&
Luu (1999). The transition between size distributions occurs at a
size of 10km.

Extrapolating the number of known KBOs of a given size results in an
estimation of 70,000 objects with sizes greater than 100km
\cite{jewitt}. A similar extrapolation for small objects, following
a size distribution of q=3.5, indicates there may be several orders
of magnitudes more objects with sizes 1km \cite{roques}.  These
numerous small objects could contain the majority of mass in the
Kuiper Belt.

\section*{Diffraction Effects}
\indent \indent Occultation of a source by an object with sharp
edges will result in diffraction effects if each point on a
resulting wavefront is considered as the center of a secondary
disturbance, giving rise to spherical wavelets.  Interference
between wavelets then causes the observed diffraction effects.  The
characteristic scale of the diffraction effects is defined as the
Fresnel scale, given by:
\begin{equation}
F=\sqrt{(\lambda D/2)}
\end{equation}
This is approximately the factor by which the width of the
occultation dip is increased.  Fresnel diffraction effects will
become important when the size of the occulting KBO is comparable
to, or larger than, that of the Fresnel scale.  At 40AU, observing
at 550nm, the Fresnel scale is 1.3km.  The characteristic length,
$\rho$, is defined as:
\begin{equation}
\rho=\frac{r}{F}
\end{equation}
For an object much smaller than the Fresnel scale ($\rho$$\ll$1),
the diffraction effects (Fig 2a) are of low amplitude and will be
obscured by noise associated with the observations. For
$\rho$$\sim$1, diffraction fringes will be observed (Fig 2b) and
increase the occultation shadow of the event. The central peak
(Poisson's Peak) is observed when the occulting object is directly
between the source ad the observer.  This effect occurs only for
spherical objects with smooth edges; KBOs will not have such
geometries.  Figure 2c shows the lightcurve for an object much
larger than the Fresnel scale ($\rho$$\gg$1). Diffraction fringes
still occur. The lightcurves presented assume a smooth, spherical
objects passes in front of a monochromatic point source.  An object
with rough edges will causes irregular diffraction fringes. In some
cases, these may be detected if the object is comparable in size to
the Fresnel scale.  The extent of diffraction is dependent on the
wavelength observed and so, in reality, the multi-wavelength flux
will cause the diffraction fringes to become smoothed.  Using a
filter reduces this effect.

\section*{Using Stellar Occultations to Detect Small KBOs}
\indent \indent The lightcurves illustrated assume the star behaves
like a point source.  In fact, the angular diameter of the star
projected at the distance of the occulting object may be larger than
the angular diameter of the object.  This would result in smoothing
of the diffraction pattern, making it harder to discern.  Thus, it
is necessary to choose target stars carefully, choosing those with
small angular diameters whilst being bright enough for fast
photometry.  The angular size of a star projected at a given
distance depends upon the spectral type and distance to the star.
The absolute magnitude and radius of a star are determined by its
mass.  The relation between spectral type and mass for main sequence
stars is used to estimate the intrinsic properties of a star.  For a
star of constant apparent magnitude, early type stars have small
angular sizes making them favorable as target stars.  A 12$^{th}$
magnitude O5V star has a projected angular size of 3$\mu$as while an
M5V star of the same magnitude will project a diameter of
280$\mu$as.

In addition to being early type on the main sequence, the stars must
be located near the ecliptic.  The classical KBO population is
concentrated within a few degrees of the ecliptic and so to maximize
the occultation rate I will restrict my search to this region. Early
type stars are typically found in the galactic plane, near their
formation regions in the spiral arms, and so the optimal region to
examine lies where the ecliptic crosses the galactic plane.

The duration of the occultation event is the ratio of the size of
the diffraction shadow to the object's apparent velocity.  When the
target star is small compared to the occulting objects the
diffraction shadow is 3 times the Fresnel scale \cite{roques2}. The
 apparent velocity of a KBO with respect to the occulted star is:
\begin{equation}
v=v_{E}[cos\omega-D^{-1/2}]
\end{equation}
where $v_{E}$ is Earth's velocity, $\omega$ is the angle from
opposition to the KBO and D is the distance (in AU) from Earth to
the KBO.  At opposition the KBO's velocity is greatest, the event
duration is shortest but the occultation rate (number of KBOs
passing in front of the star each night) is lowest.  Thus, if the
instrument used is sufficiently fast to detect dips in flux on the
timescales of the even t duration one should observe at opposition
to maximize the number of events detected.  The timescales for
events are typically less than one second, so a readout frequency on
the order of 10s of Hz is needed.

\section*{The Observations}
\indent \indent The Orthogonal Parallel Transfer Imaging Camera
mounted on the University of Hawaii 2.2m telescope on Mauna Kea will
be used to simultaneously monitor several stars.  The instrument is
designed for high-precision photometric observations and is capable
of fast readout (up to 100Hz; \cite{howell}).  The star will be
positioned in one of four 'guiding regions' near an amplifier. Only
a small designated region, $\sim$30x30 pixels, encompassing the star
will be read out. The time-varying flux from each star will be
analyzed, looking for dips greater than 3$\sigma$.  Given the short
exposure times, the target stars must be brighter than 12$^{th}$
magnitude.

By monitoring multiple stars simultaneously occultation events
caused by nearby objects (e.g. birds, planes) can be easily
identified  and discarded.  Using OPTIC, up to 4 stars can be
monitored at a time, although the availability of suitable target
stars may reduce this number when observing.  At least two stars wil
be observed per pointing to protect against spurious occultations.
Instrument noise is not expected to dominate the total noise in the
observations.  It is not necessary to measure the sky background as
this does not vary on such short timescales, instead it adds a
small, constant offset to the intensity level of the target star.
Atmospheric scintillation is the most significant source of error
for these observations.  It will be possible to discount dips caused
by scintillation through careful analysis of positive variations
from the mean and computing the standard deviation of the reading.
Only dips greater than 3$\sigma$ will be considered as possible
occultation events. The identified lightcurves can then  be more
closely examined for diffraction fringes.

OPTIC has enhanced efficiency at red wavelengths, peaking at about
750nm.  The target early type stars are generally blue, however, and
most possible target stars only have reported B and V magnitudes,
which are needed to estimate the star's projected radius at 40AU.
Futhermore, diffraction patterns are smoothed and harder to detect
at longer wavelengths.  I choose therefore to observe using a V-band
filter.

Roques \& Moncuquet \cite{roques} estimate an occultation rate of a
few occultations per night (8 hours) based on an extrapolation of
the size distribution and the number of known objects of a given
size.  I expect to detect between a few to tens of events over the
five nights of observing time.  The detectability of events depends
on the duration and the size of the occulting object.  The shortest
events will not be temporally resolved by OPTIC.  Additionally,
objects much smaller than the star's projected size at that distance
will cause only a partial reduction in flux from the star, which may
be obscured by noise.  These observations will only be sensitive to
objects larger than $\sim$1km.  Events lasting longer than a few
seconds, caused by large KBOs, will cause the instrument to lose
guiding.  These events are very unlikely but set the upper size
limit of detectable KBOs at $\sim$40km.

\section*{Summary}
\indent \indent I have presented here the case for using stellar
occultations as a method to detect small KBOs that are too faint to
observe directly. The method requires rapid photometry ($>$10Hz) on
multiple stars.  The target stars should project a size comparable
to, or smaller than, the Fresnel scale at 40AU and so early type
main sequence stars will be observed.

The planned observations will be sensitive to objects smaller than
10km.  Diffraction fringes will be observable if their amplitude is
several times that of the typical variations caused by
scintillation.  This is likely to occur for objects similar in size,
or larger than, the Fresnel scale, although scintillation will vary
from night to night.

Reported occultation events indicate the potential for this method to increase knowledge of the mass and structure of the Kuiper Belt.
The majority of mass in the outer Solar System could be contained in objects undetectable by direct methods.  Further observations will
lead to a larger, more statistically significant, sample of sub-km objects that would otherwise remain undetected.

\newpage

\textbf{Figure 1.} Evolution of the size distribution for objects in
the Kuiper Belt, assuming an intial proto-planetary disk with a mass
of 10$M_{E}$, modeled over 37Myr.  The rollover between q=3.5 and
q=4 is seen at r$\sim$10km.  The observations discussed here will be
sensitive to objects below this critical radius
\cite{kenyon}.\vspace{10ex}

\textbf{Figure 2.} Lightcurves for simulated occultations of a
monochromatic point source by a smooth, spherical object normalized
to the intensity of the target star. (a): An object much smaller
than the Fresnel scale causes low amplitude diffraction fringes.
(b): Fresnel diffraction is an important effect for objects with
$\rho\sim$1. (c): Occultation by a large object will result in total
extinction of flux from the target star.  Lightcurves simulated
using Fresnel Diffraction Explorer (http://daugerresearch.com); for
algorithm details see \cite{dauger}.\vspace{10ex}

Figures are available on YSC home page
(http://ysc.kiev.ua/abs/proc14$\_$17.pdf).

\end{document}